\documentclass[aps,prl,twocolumn,showpacs]{revtex4-1}
\usepackage{amssymb}

\usepackage{color}
\usepackage{graphicx}
\usepackage{cancel}
\usepackage{bm,amsmath}
\usepackage{dcolumn}

\def\fv{{\bf f}}

\def\xv{{\bf x}}

\def\jv{{\bf j}}

\def\vv{{\bf v}}
\def\Av{{\bf A}}

\def\Fv{{\bf F}}
\def\Jv{{\bf J}}

\def\Rv{{\bf R}}

\def\Xv{{\bf X}}

\def\lambdav{\bm{\lambda}}

\def\Im{{\rm Im}}

\begin{document}

\title{Time-dependent density functional theory for many-electron systems interacting with cavity photons}
\author{I. V. Tokatly}
\email{ilya.tokatly@ehu.es}
\affiliation{Nano-bio Spectroscopy group and ETSF Scientific
Development Centre, Departamento de F\'isica de Materiales, Universidad del Pa\'is
Vasco UPV/EHU, E-20018 San Sebast\'ian, Spain} 
\affiliation{IKERBASQUE, Basque Foundation for Science, 48011, Bilbao, Spain}

\begin{abstract}
Time-dependent (current) density functional theory for many-electron systems strongly coupled to quantized electromagnetic modes of a microcavity is proposed. It is shown that the electron-photon wave function is a unique functional of the electronic (current) density and the expectation values of photonic coordinates. The Kohn-Sham system is constructed, which allows to calculate the above basic variables by solving selfconsistent equations for noninteracting particles. We suggest possible approximations for the exchange-correlation potentials and discuss implications of this approach for the theory of open quantum systems.  
\end{abstract}

\maketitle

Time-dependent density functional theory (TDDFT) is a theoretical framework which, similarly to the ground state DFT \cite{HohKohn1964}, relies on the one-to-one mapping of the density of particles to the external potential \cite{RunGro1984}. The unique density-potential correspondence implies a possibility to calculate the exact time-dependent density by solving Hartree-like equations for fictitious noninteracting Kohn-Sham (KS) particles. This tremendous simplification of the problem makes TDDFT one of the most popular {\em ab initio} approaches for describing quantum dynamics of realistic many-body systems \cite{TDDFT-2012,TDDFTbyUllrich}.

Standard TDDFT is formulated for systems of quantum particles driven by classical electromagnetic fields \cite{RunGro1984,TDDFT-2012,TDDFTbyUllrich}, which covers most traditional problems in physics and chemistry. However, nowadays the experimental situation is rapidly changing. Progress in the fields of cavity and circuit quantum electrodynamics (QED) opens a possibility to study many-electron systems strongly interacting with quantum light. Notable examples are atoms in optical cavities in cavity-QED \cite{MabDoh2002,RaiBruHar2001,Walter2006}, or mesoscopic systems such as superconducting qubits \cite{Blais2004,Wallraff2004,YouNor2011} and quantum dots \cite{Blais2004,Wallraff2004,YouNor2011} in circuit-QED. Recently a strong coupling of molecular states to microcavity photons and the modification of chemical landscapes by cavity vacuum fields have been reported \cite{Schwartz2011,Hutchison2012,MorSte2012}. Obviously, the classical treatment of external fields prevents application of TDDFT to this new interesting class of problems.    

This paper presents TDDFT for systems of electrons strongly coupled to (or driven by) a quantized electromagnetic field \cite{Michael-note}. We prove the generalized mapping theorems for TDDFT and for time-dependent current density functional theory (TDCDFT). In both cases we analyze the structure and properties of the exchange correlation (xc) potentials and discuss possible approximation strategies. Finally we make a connection of the present theory to TDDFT for open quantum systems. 

Consider a system of $N$ electrons, e.~g., an atom or a molecule, placed inside a cavity hosting $M$ photon modes. In the Schr\"odinger picture the configuration of the system is specified by the positions $\{\xv_j\}_{j=1}^N$ of the electrons and the set $\{q_{\alpha}\}_{\alpha=1}^M$ of photonic coordinates. The full system is described by the wave function $\Psi(\{\xv_j\},\{q_{\alpha}\},t)$. Assuming as usual \cite{MultiphotonbyFaisal} that the wavelength of relevant photon modes is much larger than the size of the electronic system we adopt the dipole approximation for the electron-photon coupling. The Hamiltonian of the system takes the following form
\begin{eqnarray}
 \nonumber
 &&\hat{H} = \sum_{j=1}^N\frac{1}{2m}\Big[i\nabla_j + \Av_{\rm ext}(\xv_j,t) + 
 \sum_{\alpha}\lambdav_{\alpha}q_{\alpha}\Big]^2 \\
 &+& \sum_{i>j}W_{\xv_i-\xv_j}
 + \sum_{\alpha=1}^M\left[-\frac{1}{2}\partial_{q_{\alpha}}^2 +\frac{1}{2}\omega_{\alpha}^2q_{\alpha}^2
 - J_{\rm ext}^{\alpha}(t)q_{\alpha}\right]
 \label{H-1}
\end{eqnarray}
where $W_{\xv_i-\xv_j}$ is the electron-electron interaction, $\omega_{\alpha}$ are the frequencies of the photon modes, and $\lambdav_{\alpha}$ describes coupling to the $\alpha$-mode. The electron and photon subsystems can be driven externally by the classical vector potential $\Av_{\rm ext}(\xv,t)$ and the external ``currents`` $J_{\rm ext}^{\alpha}(t)$. The vector potential describes (in a temporal gauge) forces from the ions in atoms and molecules and all other possible classical fields. The currents $J_{\rm ext}^{\alpha}(t)$ allow for an external excitation of the cavity modes. The time evolution from a given initial state $\Psi_0(\{\xv_j\},\{q_{\alpha}\})$ is governed by the Schr\"odinger equation
\begin{equation}
 \label{SE-1}
 i\partial_t \Psi(\{\xv_j\},\{q_{\alpha}\},t) = \hat{H}\Psi(\{\xv_j\},\{q_{\alpha}\},t).
\end{equation}
Solution of this equation gives a complete description of the system for a fixed configuration of the external fields. 

All DFT-like approaches assume that the state of the system is also uniquely determined by a small set of basic observables, such the density in TDDFT, the current in TDCDFT, and possibly something else in other generalizations of the theory. Below we construct two generalizations of TDDFT for electron-photon systems described by the Hamiltonian of Eq.~(\ref{H-1}).

Let us start with a technically simpler current-based theory. Consider the electronic current $\jv(\xv,t)$ and expectation values $Q_{\alpha}(t)$ of photon coordinates as the basic variables. These variables are defined as follows
\begin{eqnarray}
 \label{Q-def}
 &&Q_{\alpha} = \langle\Psi|q_{\alpha}|\Psi\rangle, \\
 \label{j-def}
 &&\jv = \langle\Psi|\hat{\jv}_p(\xv)|\Psi\rangle - \frac{n}{m}\Av_{\rm ext} 
 - \sum_{\alpha}\frac{\lambdav_{\alpha}}{m}\langle\Psi|q_{\alpha}\hat{n}(\xv)|\Psi\rangle,
 \end{eqnarray}
where $n(\xv,t)=\langle\Psi|\hat{n}(\xv)|\Psi\rangle$ is the electronic density, and $\hat{n}(\xv)=\sum_{j}\delta(\xv - \xv_j)$ and $\hat{\jv}_p(\xv)= \frac{-i}{2m}\sum_{j}\{\nabla_j,\delta(\xv - \xv_j)\}$ are the density and paramagnetic current operators.

Equations of motion for $Q_{\alpha}$ follow Eqs.~(\ref{Q-def}) and (\ref{SE-1}):
\begin{equation}
 \label{Q-EOM}
 \ddot{Q}_{\alpha} + \omega_{\alpha}^2Q_{\alpha} = \lambdav_{\alpha}\Jv(t) + J_{\rm ext}^{\alpha}(t)
\end{equation}
where $\Jv(t)=\int\jv(\xv,t)d\xv$ is the space-averaged electronic current. Equation~(\ref{Q-EOM}) is simply the Maxwell equation for the cavity vector potential projected on the $\alpha$-mode.

Equations (\ref{H-1})--(\ref{j-def}) determine the wave function $\Psi(t)$, and the basic variables $\jv$ and $Q_{\alpha}$ as functionals of the initial state $\Psi_0$, and the external fields $\Av_{\rm ext}$ and $J_{\rm ext}^{\alpha}$. This defines a unique map $\{\Psi_0,\Av_{\rm ext},J_{\rm ext}^{\alpha}\} \mapsto\{\Psi,\jv,Q_{\alpha}\}$. TDCDFT assumes the existence of a unique ''inverse map'' $\{\Psi_0,\jv,Q_{\alpha}\}\mapsto\{\Psi,\Av_{\rm ext},J_{\rm ext}^{\alpha}\}$. That is, given the initial state and the basic observables one can uniquely recover the full wave function and the external fields that generate the prescribed dynamics of the basic variables. 
 
To prove the uniqueness of the inverse map we follow the nonlinear Schr\"odinger equation (NLSE) approach \cite{TokatlyChemPhys2011,Maitra2010}. Assume that $\jv(\xv,t)$ and $Q_{\alpha}(t)$ are given, and express the external fields from Eqs.~(\ref{j-def}) and (\ref{Q-EOM}) as follows,
\begin{eqnarray}
 \label{Aext}
 \Av_{\rm ext} &=& \frac{m}{n}\langle\Psi|\hat{\jv}_p(\xv)-\jv|\Psi\rangle 
 - \sum_{\alpha}\frac{\lambdav_{\alpha}}{n}\langle\Psi|q_{\alpha}\hat{n}(\xv)|\Psi\rangle, \\
 \label{Jext}
 J_{\rm ext}^{\alpha} &=& \ddot{Q}_{\alpha} + \omega_{\alpha}^2Q_{\alpha} - \lambdav_{\alpha}\Jv.
\end{eqnarray}
This defines the external fields as explicit functionals of the observables $\jv(\xv,t)$ and $Q_{\alpha}(t)$, and the instantaneous state $\Psi(t)$. Substitution of Eqs.~(\ref{Aext}) and (\ref{Jext}) into the Hamiltonian~(\ref{H-1}) turns Eq.~(\ref{SE-1}) into the many-body NLSE
\begin{equation}
 \label{NLSE-1}
 i\partial_t \Psi(t) = \hat{H}[\jv,Q_{\alpha},\Psi]\Psi(t),
\end{equation}
where $\hat{H}[\jv,Q_{\alpha},\Psi]$ is an instantaneous functional of $\Psi(t)$, which depends parametrically on $\jv(\xv,t)$ and $Q_{\alpha}(t)$. The uniqueness of a solution to Eq.~(\ref{NLSE-1}) can be proven easily under the usual in TD(C)DFT assumption of $t$-analyticity \cite{RunGro1984,vanLeeuwen1999,Vignale2008}. Assuming that $\jv(\xv,t)$ and 
$Q_{\alpha}(t)$ are analytic functions in time, we represent them and unknown $\Psi(t)$, by the Taylor series
$$
\jv(t)=\sum_{k=0}^{\infty}\jv^{(k)}t^k, \,\, Q_{\alpha}(t)=\sum_{k=0}^{\infty}Q_{\alpha}^{(k)}t^k,
 \,\, \Psi(t) =\sum_{k=0}^{\infty}\Psi^{(k)}t^k.
$$
After inserting these series into Eq.~(\ref{NLSE-1}) one observes that all coefficients $\Psi^{(k)}$ with $k>0$ can be expressed recursively in terms of $\jv^{(k)}$, $Q_{\alpha}^{(k)}$, and $\Psi^{(0)}\equiv\Psi_0$. The simple reason for this is that the right hand side in Eq.~(\ref{NLSE-1}) is an instantaneous functional of $\Psi(t)$, while the left hand side $\sim\partial_t\Psi(t)$. As the recursion produces the unique Taylor series for $\Psi(t)$, the many-body wave function is a unique functional of the initial state and the basic variables, $\Psi[\Psi_0,\jv,Q_{\alpha}]$. By substituting this wave function into Eq.~(\ref{Aext}) we find the functional $\Av_{\rm ext}[\Psi_0,\jv,Q_{\alpha}]$, which completes the proof of the TDCDFT mapping theorem.

The KS system for this theory is constructed as follows. Consider a system of $N$ noninteracting particles coupled to the photon modes at the mean field level. This system is described by a set of $N$ one-particle KS orbitals $\varphi_j(\xv,t)$ which satisfy the following equations
\begin{equation}
 \label{KSeq-1}
 i\partial_t\varphi_j = 
 \frac{1}{2m}\Big(i\nabla + \Av_S + \sum_{\alpha}\lambdav_{\alpha}Q_{\alpha}\Big)^2\varphi_j,
\end{equation}
where $Q_{\alpha}(t)$ is the solution to Eq.~(\ref{Q-EOM}) with $\Jv(t)$ being replaced by the space-average of the KS current density
\begin{equation}
 \label{jKS}
 \jv_{S} = \frac{1}{m}\sum_j\Im(\varphi_j^*\nabla\varphi_j) 
 - \frac{n}{m}\Big(\Av_S + \sum_{\alpha}\lambdav_{\alpha}Q_{\alpha}\Big).
\end{equation}

Using the above NLSE argumentation (or the standard TDCDFT mapping \cite{Vignale2008}) we find that $\varphi_j(\xv,t)$ are unique functionals of the KS current $\jv_{S}(\xv,t)$ and the KS initial state $\Psi_{0}^{S}$. A comparison of Eqs.~(\ref{j-def}) and (\ref{jKS}) shows that the KS current reproduces the physical current $\jv_S=\jv$ if $\Av_{S}$ in Eq.~(\ref{KSeq-1}) is defined as $\Av_S=\Av_{\rm ext} + \Av_{\rm Hxc}$, where
\begin{eqnarray}
 \nonumber
 \Av_{\rm Hxc} &=&  
 \frac{1}{n}\Big[\sum_j\Im(\varphi_j^*\nabla\varphi_j)
 -m\langle\Psi|\hat{\jv}_p(\xv)|\Psi\rangle\\
 &+&  \sum_{\alpha}\lambdav_{\alpha}\langle\Psi|\Delta q_{\alpha}\Delta\hat{n}(\xv)|\Psi\rangle\Big]
 \label{AHxc}
\end{eqnarray}
Here $\Delta q_{\alpha}=q_{\alpha}-Q_{\alpha}(t)$, and $\Delta\hat{n}(\xv)=\hat{n}(\xv)-n(\xv,t)$ are the fluctuation operators for the photonic coordinates and the electronic density, respectively. By construction, for given initial states $\Psi_0$ and $\Psi_0^S$, the potential $\Av_{\rm Hxc}$ is a gauge invariant universal functional of $\jv$ and $Q_{\alpha}$.

Therefore the current and the photonic coordinates can be calculated from a system of Eqs.~(\ref{KSeq-1}), (\ref{Q-EOM}) describing noninteracting fermions driven by a selfconsistent field and coupled to a set of classical harmonic oscillators. There is a deep reason for explicitly separating the ''mean-field'' part $\Av_{\rm mf}=\sum_{\alpha}\lambdav_{\alpha}Q_{\alpha}$ of the selfconsistent potential in Eq.~(\ref{KSeq-1}). This part accounts for the net force exerted on electrons from the photons. The remaining Hxc-part $\Av_{\rm Hxc}$ does not produce a global force,
\begin{equation}
 \label{zero-force-1}
 \int \big[\jv\times(\nabla\times\Av_{\rm Hxc}) - n\partial_t\Av_{\rm Hxc}\big] d\xv = 0,
\end{equation}
which can be checked directly using Eqs.~(\ref{AHxc}), (\ref{SE-1}), and (\ref{KSeq-1}). Apparently both Hartree and xc contributions to $\Av_{\rm Hxc}=\Av_{\rm H}+\Av_{\rm xc}$ satisfy the identity of Eq.~(\ref{zero-force-1}) independently. Equation~(\ref{zero-force-1}) is a generalization of the zero-force theorem \cite{Vignale1995a} for the considered electron-photon system. In this regard $\Av_{\rm xc}$ is similar to the xc potential in the usual TDCDFT for closed purely electronic systems.

If the electrons are driven by a scalar external potential $V_{\rm ext}(\xv,t)$, one can choose the density $n(\xv,t)$ as the basic variable for electronic degrees of freedom. Let us identify the photonic basic observables for the electron-photon TDDFT. First, we transform of the photon field in Eq.~(\ref{H-1}) from the velocity to the length gauge \cite{MultiphotonbyFaisal}. Then we perform the canonical transformation of photon variables, $i\partial_{q_{\alpha}}\mapsto\omega_{\alpha}p_{\alpha}\,$, $\, q_{\alpha}\mapsto -i\omega_{\alpha}^{-1}\partial_{p_{\alpha}}$, which ``exchanges'' the photon coordinates and momenta while preserving their commutation relations. The system is now described by the wave function 
$\Phi(\{\xv_j\},\{p_{\alpha}\},t)$ that is the Fourier transform of $\Psi(\{\xv_j\},\{q_{\alpha}\},t)$. 
The final transformed Hamiltonian takes the form
\begin{eqnarray}
 \nonumber
 &&\hat{H} = \sum_j\left[-\frac{\nabla_j^2}{2m} + V_{\rm ext}(\xv_j,t) \right]+ \sum_{i>j}W_{\xv_i-\xv_j}\\
 &&+ \sum_{\alpha}\Big[-\frac{1}{2} \partial_{p_{\alpha}}^2
 +\frac{\omega_{\alpha}^2}{2}\Big(p_{\alpha} - \frac{\lambdav_{\alpha}}{\omega_{\alpha}}\hat{\Xv} \Big)^2
 + \frac{\dot{J}_{\rm ext}^{\alpha}(t)}{\omega_{\alpha}}p_{\alpha}\Big], \,\,
 \label{H-2}
\end{eqnarray}
where $\hat{\Xv}=\sum_{j=1}^N\xv_j$.  
The structure of Eq.~(\ref{H-2}) suggests that the proper basic variables are the density  $n(\xv,t)$ and the expectation values $P_{\alpha}(t)$ of the photon momenta
\begin{equation}
 \label{P-def}
 n(\xv,t)=\langle\Phi|\hat{n}(\xv)|\Phi\rangle, \,\, P_{\alpha}(t) = \langle\Phi|p_{\alpha}|\Phi\rangle.
\end{equation}

Equations of motion for the basic variables read
\begin{eqnarray}
 \label{P-EOM}
 &&\ddot{P}_{\alpha} + \omega_{\alpha}^2P_{\alpha} - \omega_{\alpha}\lambdav_{\alpha}\Rv =
 - {\dot{J}_{\rm ext}^{\alpha}}/{\omega_{\alpha}}, \\
\label{n-EOM}
 && m\ddot{n} + \nabla\Fv_{\rm str} + \sum_{\alpha}\nabla\fv_{\alpha} = \nabla (n\nabla V_{\rm ext}),
\end{eqnarray}
where $\Rv(t)=\langle\Phi|\hat{\Xv}|\Phi\rangle=\int \xv n(\xv,t)d\xv$ is the expectation value of the center of mass coordinate, and $\Fv_{\rm str} = im\langle\Phi|[\hat{T}+\hat{W},\hat{j}_p]|\Phi\rangle=-\nabla\tensor{\Pi}$ is the usual electronic stress force which is equal to the divergence of the electronic stress tensor. The force 
$\fv_{\alpha}(\xv,t)$ exerted from the $\alpha$-photon mode on electrons is given by the expression
\begin{equation}
 \label{f}
 \fv_{\alpha}(\xv,t) = \lambdav_{\alpha}\langle\Phi|(\omega_{\alpha}p_{\alpha}
 -\lambdav_{\alpha}\hat{\Xv})\hat{n}(\xv)|\Phi\rangle .
\end{equation}

Now we are ready to prove the uniqueness of the map $\{\Phi_0,n,P_{\alpha}\}\mapsto\{\Phi,V_{\rm ext},J_{\rm ext}^{\alpha}\}$ from the initial state and the observables to the time-dependent wave function and the external fields.  The corresponding many-body NLSE is constructed using Eqs.~(\ref{P-EOM}) and (\ref{n-EOM}).

By solving Eqs.~(\ref{P-EOM}) and (\ref{n-EOM}) for $\dot{J}_{\rm ext}^{\alpha}$ and $V_{\rm ext}$ we get the external fields as functionals of the basic variables and the instantaneous wave function $\Phi(t)$, $V_{\rm ext}[\Phi,n]$ and $\dot{J}_{\rm ext}^{\alpha}[n,P_{\alpha}]$. Inserting these functionals into Eqs.~(\ref{H-2}) we obtain a $\Phi$-dependent Hamiltonian $\hat{H}[n,P_{\alpha},\Phi]$ of the many-body NLSE that determines the wave function for a given initial state $\Phi_0$, the density $n(t)$ and photon momenta $P_{\alpha}(t)$. The uniqueness of a solution to this NLSE is demonstrated in exactly the same way as for the above TDCDFT case, provided the standard $t$-analyticity conditions are fulfilled. This proves the generalized TDDFT mapping theorem: the many-body wave function and the external fields are the unique functionals of the basic variables, $n$ and $P_{\alpha}$, and the initial state \cite{math-note}.

The KS system can be again constructed explicitly. Consider a system of noninteracting particles described by $N$ KS orbitals which satisfy the equations
\begin{equation}
 \label{KSeq-2}
i\partial_t\phi_j = -\frac{\nabla^2}{2m}\phi_j + \Big[V_S +\sum_{\alpha}( 
\omega_{\alpha}P_{\alpha} - \lambdav_{\alpha}\Rv)\lambdav_{\alpha}\xv\Big]\phi_j,
\end{equation}
where the second term in the square brackets is the mean-field analog of the electron-photon interaction term in Eq.~(\ref{H-2}). The force balance equation for this system takes the form
\begin{equation}
 \label{nKS-EOM}
m\ddot{n} + \nabla\Fv_{\rm str}^{S} +
\nabla\sum_{\alpha}\lambdav_{\alpha}( \omega_{\alpha}P_{\alpha} - \lambdav_{\alpha}\Rv)n= \nabla (n\nabla V_{S}),
\end{equation}
where $\Fv_{\rm str}^S = im\langle\Phi^S|[\hat{T},\hat{j}_p]|\Phi^S\rangle=-\nabla\tensor{\Pi}_S$ is the kinetic stress force of noninteracting fermions [$\Phi^S(t)$ is the KS Slater determinant]. By applying the NLSE arguments to Eqs.~(\ref{KSeq-2})-(\ref{nKS-EOM}) we conclude that $\varphi_j$ and $V_S$ are unique functionals of $n(\xv,t)$, $P_{\alpha}(t)$, and the KS initial state $\Phi_0^S$. Then, from Eqs.~(\ref{n-EOM}) and (\ref{nKS-EOM}) one finds that the KS density reproduces the exact density if $V_S$ is of the form
\begin{equation}
 \label{VS}
V_S = V_{\rm ext} + V_{\rm Hxc}^{\rm el} + \sum_{\alpha}V_{\rm xc}^{\alpha},
\end{equation}
where the universal functionals $V_{\rm Hxc}^{\rm el}[n,P]$ and $V_{\rm xc}^{\alpha}[n,P]$ are defined via the following Sturm-Liouville problems
\begin{eqnarray}
\label{VHxc-el}
&& \nabla (n\nabla V_{\rm Hxc}^{\rm el}) = \nabla (\Fv_{\rm str}^S - \Fv_{\rm str})
= \nabla (\nabla\tensor{\Pi}_{\rm Hxc}),\\
&& \nabla (n\nabla V_{\rm xc}^{\alpha}) = \nabla\lambdav_{\alpha}
\langle\Phi|(\lambdav_{\alpha}\Delta \hat{\Xv} - \omega_{\alpha}\Delta p_{\alpha})\Delta \hat{n}|\Phi\rangle,
\label{Vxc-alpha}
\end{eqnarray}
with $\Delta p_{\alpha}=p_{\alpha}-P_{\alpha}(t)$ and $\Delta\hat{\Xv}=\hat{\Xv}-\Rv(t)$ being the fluctuation operators for the photon momenta and the center of mass coordinate of the electrons.
    
Interestingly, in TDDFT the total xc potential is naturally separated into the usual electronic stress contribution and the contributions assigned to each photon mode. It is obvious from Eqs.~(\ref{VHxc-el}) and (\ref{Vxc-alpha}) that each contribution to the total xc potential satisfies the zero-force theorem, $\int n\nabla V_{\rm Hxc}^{\rm el}d\xv =\int n\nabla V_{\rm xc}^{\alpha}d\xv =0$.
The net photon force exerted on electrons is fully captured by the mean field electron-photon potential in Eq.~(\ref{KSeq-2}). The zero force theorem is a consequence of the harmonic potential theorem (HPT) \cite{Dobson1994,Vignale1995a}, which also holds true here as photons form a set of harmonic oscillators coupled bilinearly to the electronic center of mass.

In practice any DFT-type approach requires approximations for xc potentials. In the present generalization of the theory we succeeded to define the xc potential in such a way that it has the same general properties, and obeys the same set of constraints as the xc potential in the usual purely electronic TDDFT. This suggests natural strategies for constructing approximations.   

(i) The first possibility is a velocity gradient expansion. At zero level we set $V_{\rm xc}^{\alpha}=0$ and take $V_{\rm xc}^{\rm el}=V_{\rm xc}^{\rm ALDA}$, the xc potential in the standard adiabatic local density approximation  (ALDA). This seemingly naive approximation exactly reproduces the correct HPT-type dynamics as for the rigid motion with a uniform velocity $\vv=\jv/n$ the effect of photons on the density dynamics is exhausted by the mean-field contribution. The zero level approximation can be viewed as a generalization of ALDA. The dynamical corrections should be proportional to velocity gradients. It should be possible to derive them perturbatively in the TDCDFT scheme along the lines of the Vignale-Kohn approximation \cite{VigKohn1996,VigUllCon1997}.

(ii) Probably a more promising strategy is to make a connection of the effective KS potential to the many-body theory \cite{GattiTokPRL2007}. Beyond the mean field level the electron-photon coupling generates a retarded photon-mediated interaction between the electrons. The corresponding photon propagators will enter the diagrams for the electronic self energy as additional interaction lines. The new contribution to the self energy can then be connected to the xc potential via Sham-Schl\"utter equation \cite{ShaSch1983,vanLeeuwen1996}. In principle the corresponding xc potential can be constructed perturbatively to any desired order in the coupling constant \cite{TokPanPRL2001}. However, already the simplest approximation, generated by the exchange-like diagram, is expected to capture the important physics. Formally this approximation for $V_{\rm xc}^{\alpha}$ is an analog of the x-only optimized effective potential \cite{UllGosGro1995,vanLeeuwen1996}. Physically it should be responsible for the Lamb shift effects and for the spontaneous photon emission in nonequilibrium situations.

Exploring practical performance of these approximations is an interesting direction for the future research.

If the functionals $V_{\rm Hxc}[n,P_{\alpha}]$ and $V_{\rm xc}^{\alpha}[n,P_{\alpha}]$ are known, the basic variables, $n(\xv,t)$ and $P_{\alpha}(t)$, can be calculated by solving Eqs.~(\ref{KSeq-2}), (\ref{P-EOM}). In general the KS Eq.~(\ref{KSeq-2}) should be solved numerically, while Eq.~(\ref{P-EOM}) always admits an analytic solution. For example, for the equilibrium initial state and $J_{\rm ext}^{\alpha}=0$ this solution reads 
\begin{equation}
 \label{P-t}
 P_{\alpha}(t) = \int_{0}^t\sin[\omega_{\alpha}(t-t')]\lambdav_{\alpha}\Rv(t')dt'.
\end{equation}
By substituting Eq.~(\ref{P-t}) into Eq.~(\ref{KSeq-2}) we eliminate the photon variables and get the KS equation involving only the electronic density. Thus we obtain TDDFT for an open quantum system -- now the KS equation describes in a closed form only the electronic part of the full electron-photon system. 

Formally Eq.~(\ref{H-2}) is a version of the Caldeira-Leggett (CL) model \cite{CalLeg1983a,CalLeg1983b}. Therefore, as a byproduct we obtained TDDFT for open systems coupled to the CL bath of harmonic oscillators. Let us assume Ohmic spectral density of the bath, 
$\pi\sum_{\alpha}\lambda_{\alpha}^{\mu}\lambda_{\alpha}^{\nu}\delta(\omega -\omega_{\alpha})=2\eta\delta^{\mu\nu}$, where $\eta$ is the friction constant. In this case the selfconsistent potential in Eq.~(\ref{KSeq-2}) reduces to the form 
$V_{\rm eff}=V_{\rm Hxc}+\eta N\dot\Rv \xv$. The last, mean field term is exactly the potential in the phenomenological dissipative NLSE proposed by Albrecht \cite{Albrecht1975,Hasse1975}. Hence already at zero level we recover one of the heuristic theories of quantum dissipation. Deficiencies of the Albrecht equation should be corrected by going beyond the zero level approximation.


Currently there are several formulations of TDDFT for open systems, based on the master equation for the density matrix \cite{BurCarGeb2005,Zhou2009,Zhou2010}, or on the stochastic Schr\"odinger equation \cite{VenAgo2007,AgoVen2008}. At the level of the final KS equations our theory is similar to the formulation of Refs.~\cite{Zhou2009,Zhou2010} which also allows for the unitary propagation of KS orbitals. The conceptual difference is that in the present case both the TDDFT mapping and the approximation strategies are universally valid for the cavity situation with a few discrete photon modes and for the bath with a continuous spectral density. The bath in traced out at the very last step after setting up the TDDFT framework together with approximations.

In conclusion, TD(C)DFT for systems strongly coupled to the cavity photon fields is proposed. We proved the corresponding generalizations of the mapping theorems, established the existence of the KS system, and suggested a few technically feasible approximation strategies. In the limit of dense spectrum of photon modes this approach naturally leads to TD(C)DFT for open quantum systems. This work is a step towards {\em ab initio} theory of various cavity/circuit QED experiments, and practical TDDFT for dissipative systems. 

This work was supported by the Spanish MEC (FIS2007-65702-C02-01). 


\begin{thebibliography}{10}
\bibitem{math-note}%
  \BibitemOpen
  \bibinfo {note} {The level of mathematical rigor and the conditions of the
  present mapping theorems are exactly the same as for the standard purely
  electronic TD(C)DFT. Strictly speaking, no extra assumption is required if
  the number of photon modes is arbitrary large, but finite.}%
  \bibAnnoteFile{Stop}{math-note}%
\end{thebibliography}

%

\end{document}